\documentclass[]{article}
\usepackage[latin5]{inputenc}
\usepackage {amssymb}
\usepackage {amsmath}
\usepackage{time}

\newcommand{\be}{\begin{equation}}
\newcommand{\ee}{\end{equation}}
\newcommand{\beq}{\begin{eqnarray}}
\newcommand{\eeq}{\end{eqnarray}}

\begin{document}

\title{Another  Model  with Interacting Composites}

\author{M. Hortaçsu$^{*\dag}$ and F. Taşkın$^{*\ddag}$ \\%
\small$^*$Department of Physics, Istanbul Technical University, Istanbul,34469 Turkey\\%
\small$^\dag$Feza G\"{u}rsey Institute, Istanbul, Turkey\\%
\small$^\ddag$Department of Physics, Erciyes University, Kayseri,
38090 Turkey}

\maketitle

\begin{abstract}
We show that we can construct
 a model  in $3+1$ dimensions where it is necessary that
 composite vector particles take place in physical processes as
incoming and outgoing particles . Cross-section of the processes
in which only the constituent spinors take place goes to zero.
While the spinor-spinor scattering goes to zero, the scattering of
composites gives nontrivial results.
\end{abstract}

\section{Introduction}
To find a nontrivial field theoretical model is one of the
outstanding problems in theoretical high energy physics.  The
perturbatively nontrivial $\phi^4$ theory in four dimensions was
shown to go to a free theory as the cut-off is lifted
\cite{baker,wilson}.  The Nambu-Jona-Lasinio model \cite{nambu},
which is non-renormalizable perturbatively, was also shown to go
to a trivial model \cite{kogut,Zinn-Justin}.  There are claims
that the gauged Nambu-Jona-Lasinio model \cite{bardeen1,bardeen2}
may give a non-trivial theory for a non realistically large number
of flavors \cite{Reenders1,Reenders2}. All these examples show
that the search for non-trivial models may be an interesting
endeavor.

On the other hand, there are two-dimensional models with infinite
number of local and non-local charges \cite{Brezin,Eichenherr}.
These models were shown to give scattering matrices without
particle production \cite{Zamo,Weisz} indicating a behavior not
expected from fully interacting theories.  When the symmetry
present in these models is broken, these models give rise to
particle production in perturbative calculations \cite{thun}.

Another class of models are those with zero scattering amplitudes
for the constituent fields, even though their Lagrangians seem to
have non-zero interaction terms. These interactions arise from
constraints.  An example of them is taking a product of the
constituent field equal to a power of the auxiliary field
\cite{ak1,ak2,ak3}, i.e. forming composites of the constituent
fields. The interaction of the constituent field with the
composite field is defined. The Lagrangian contains the kinetic
part of only the constituent spinor fields and the kinetic terms
for the composites are formed as a result of vacuum polarizations
of the composite field due to its interaction with the constituent
field. Such models simulate the models with only spinors.

The first work on models with only spinors goes back to the work
of Heisenberg\cite{heisenberg}. Gursey proposed his model as a
substitute for the Heisenberg model in fifties \cite{gursey}. This
spinor model is important since it is conformally invariant
classically and has classical solutions \cite{kortel} which may be
interpreted as instantons and merons \cite{ak}, similar to the
solutions of pure Yang-Mills theories in four dimensions.  This
original model can be generalized to include vector, pseudovector
and pseudoscalar interactions.

A while ago, one of us, M.H.,with collaborators, claimed to have
written a polynomial Lagrangian equivalent  to the one given by
Gursey \cite{ak1,ak2,ak3}. Now we know that the models studied are
only naively equivalent to these spinor models.

He, with collaborators, has also shown that
\cite{ak1,ak2,ak3,arik} in some of these models, the interaction
of the spinor field with other spinors goes to zero. The model is
not only asymptotically free, it is actually free, meaning that
the coupling constant goes to zero as the cut-off is removed. We
could not find physical processes \cite{arik} which give results
indicative of an interacting theory.

There is one more aspect of these models we missed in our work in
the eighties \cite{ak1,ak2,ak3,arik}. A further study shows that
although the constituent fields do not give finite scattering
amplitudes with each other, two composites can scatter from each
other and may give rise to composite particle creation with finite
results. These fact makes these models, which we had discarded as
"trivial", interesting once more. M.H., with a collaborator, has
already studied a model of this class with scalar composites
\cite{hortacsucan}. In this work we will study another form of
these models.  We will first show why the constraint model is not
totally equivalent to the spinor model with non polynomial
interaction.  Then we will repeat the calculations of  given in
reference  \cite{ak2}. Higher order calculations and the
calculation of processes where the composites take place will be
studied in the next section. Since the two models are similar,
there will be some repetition of the work in \cite{hortacsucan}.
We will end with remarks where this model is coupled to a
constituent scalar field. This is a model where  the scalar field
can be considered as coupled to the spinor field via the gauge
field, with the difference from the usual case that here the gauge
field is the induced composite field.

\section{An additional symmetry}

We start with the pure spinor Lagrangian

 \be L=\overline{\psi}\left(i\partial\!\!\!/ -ieg\partial
   \!\!\!/g^{-1}-m\right)\psi+\alpha \left[(\overline{\psi}
   \gamma_{\mu}\psi)(\overline{\psi}\gamma^{\mu}\psi)\right]^{2/3} +
   s\left[\overline{w}^{\mu}(\overline{\psi}\gamma_{\mu}\psi-G_{\mu}G^2)
   \right]. \label{gursey lagrangian}\ee

This is the vector form of the Gursey model \cite{gursey}. Here
only the spinor fields have their kinetic part. The $g$ field is
used to construct a pure gauge term, just to restore the local
gauge symmetry, when the spinor field is transformed. We introduce
an auxiliary field $G^{\mu}$ and a anti ghost field
$\overline{w}_{\mu} $ into the Lagrangian and also add a term
using $s$ as a symmetry operator to give us the constraint and
Faddeev-Popov terms.  The symmetry $s$ transforms the fields as

 \beq
   s\bar {\omega}^{\mu} = \lambda^{\mu},\hspace{5mm} s\lambda^{\mu} = 0,
   \hspace{5mm}sG^{\mu} = \omega^{\mu} \nonumber \\
   s\omega^{\mu}=0, \hspace{5mm} s\psi = s\bar \psi = 0 \hspace{5mm}
   s^2=0\label{s donusum}
 \eeq

The $g$ field also transforms trivially.

Our Lagrangian should be invariant under this operation.
Performing the $s$ operation, we rewrite our Lagrangian as

\beq L=\overline{\psi}\left(i\partial\!\!\!/ -ieg\partial
 \!\!\!/g^{-1}-m\right)\psi+\alpha \left[\overline{\psi}
 \gamma_{\mu}\psi\overline{\psi}\gamma^{\mu}\psi\right]^{2/3}  \nonumber \\
 +\lambda^{\mu}\left(\overline{\psi}\gamma_{\mu}\psi-G_{\mu}G^2\right)+
 \overline{w}^{\mu}(G^2+2w_{\lambda}G^{\lambda})w_{\mu}.\eeq

Since the spinor field vanishes under the $s$ operation we do not
get any contribution from this part.  Acting with $s$ on the total
Lagrangian gives

\beq
 s L=-\lambda^{\mu}w_{\mu}G^2-2\lambda^{\mu}G_{\mu}w_{\lambda}G^{\lambda}
 +\lambda^{\mu}w_{\mu}G^2+2\lambda^{\mu}G_{\mu}w_{\lambda}G^{\lambda}\nonumber \\
 +2\overline{w}^{\mu}w_{\mu}w_{\lambda}G^{\lambda}-2\overline{w}^{\mu}w_{\mu}w_{\lambda}G^{\lambda}
 +2\lambda^{\mu}G_{\mu}w_{\lambda}w^{\lambda}=0 \eeq

This result  validates our assertion.

In a paper written by M.H.,  with collaborators, \cite{ak3} it has
been asserted that the model given by the above Lagrangian is
equivalent with the model described by the Lagrangian given below.

 \be \acute{L}=\overline{\psi}\left(i\partial \!\!\!/ -ieg\partial
   \!\!\!/g^{-1}+eG_{\mu}\gamma^{\mu}+e\lambda_{\mu}\gamma^{\mu}-m\right)\psi
   -e^4\lambda_{\mu}G^{\mu}\left(G_{\lambda}G^{\lambda}\right)+\mbox{ghost
terms} \label{s yenilagran} \ee

Although the new model has the correct relation between the
auxiliary field $G^{\mu}$ and the spinor field, we see that
replacing the fractional spinor interaction by a spinor vector
coupling may not be allowed.  The pure spinor term with fractional
powers have the symmetry described by the $s$ operation, whereas
the latter term does not.  From this point on we will study the
properties of   the second model.  We will view the Gursey model
as a motivation, a pure spinor model which is only naively
equivalent to the latter model.

We will find that although the $G^{\mu}$ field does not have a
kinetic term in the original Lagrangian, the one-loop corrections
will generate them, making this composite field behave as a
dynamical entity.  In the literature there are similar models with
differential operators in the interaction Lagrangian \cite{amati}.
We will not take such terms in our model.

\section{The Model}

We start with the polynomial Lagrangian, where  the fractional
interaction of eq.(\ref{gursey lagrangian}) is replaced by the
product of fields and a constraint term.

  \be L=\overline{\psi}\left(i\partial \!\!\!/ -ieg\partial \!\!\!/
    g^{-1}+eG_{\mu}\gamma^{\mu}+e\lambda_{\mu}\gamma^{\mu}-m\right)\psi
    -e^4\lambda_{\mu}G^{\mu}\left(G_{\lambda}G^{\lambda}\right)
    \label{yenilagran} \ee

In this expression only the spinor fields have the kinetic part.
$\lambda_{\mu}$ and $G_{\mu}$ are two  auxiliary vector fields. If
we write the Euler-Lagrange equations for the $\lambda_{\mu}$ and
$G_{\mu}$ fields, the equations of motion give

 \beq
 \lambda_{\mu}(\overline{\psi}\gamma^{\mu}\psi-e^3G^{\mu}G^2)&=&0, \nonumber \\
 \overline{\psi}\gamma^{\mu}\psi-e^3(\lambda^{\mu}G^{\kappa}G_{\kappa}+2\lambda^{\kappa} G_{\kappa} G^{\mu})&=&0. \eeq

Since we have a constraint Lagrangian, we have to perform the
constraint analysis $\grave{a}$ la Dirac \cite{dirac}.

First we have the spinor-Dirac constraints. The auxiliary fields
give us the extra constraints

\beq
\Sigma^{\mu}&=&\delta L/\delta(\partial_{0}\lambda_{\mu})\approx 0 \nonumber \\
\Omega^{\mu}&=&\delta L/\delta(\partial_{0}G_{\mu})\approx 0 \nonumber \\
 \eeq

To write the canonical hamiltonian we use the relation

 \be H_{c}=p_{i}\dot{q_{i}}-L \ee

which gives

 \be H_{c}=\overline{\psi}\left(i\gamma^{i}\partial_i+eg\partial
  \!\!\!/ g^{-1} - eG_{\mu}\gamma^{\mu} - e\lambda_{\mu}\gamma^{\mu}
  + m\right)\psi + e^4\lambda_{\mu}G^{\mu}
  \left(G_{\lambda}G^{\lambda}\right) \ee

The constraints are added to the canonical expression to give

 \beq  H_{p}=\overline{\psi}\left(i\gamma^{i}\partial_i +eg\partial
  \!\!\!/ g^{-1} - eG_{\mu}\gamma^{\mu} - e\lambda_{\mu}\gamma^{\mu}
  + m\right)\psi + e^4\lambda_{\mu}G^{\mu}
  \left(G_{\lambda}G^{\lambda}\right) \nonumber \\+ a\pi+
  (\overline{\pi}-i\overline{\psi}\gamma_{0})b+c_{\mu}\Sigma^{\mu}+d_{\mu}\Omega^{\mu}
  \hspace{30mm} \eeq

Here a, b, $c_{\mu}, d_{\mu}$   are Lagrange multipliers. The
condition that the constraints should not change in time dictates
that the Poisson brackets of the constraints with the Hamiltonian
must vanish.

\be \{\theta_{i},H_{p}\}=0 \ee

$a$ and $b$   are evaluated by these relations.

The rest of the Poisson parentheses give us further constraints.

\beq
\kappa^{\mu}&=&\overline{\psi}\gamma^{\mu}\psi-G^{\nu}G_{\nu} G^{\mu} \nonumber \\
\zeta^{\mu}&=&G^{\nu}
G_{\nu}G^{\mu}-\overline{\psi}\gamma^{\mu}\psi-2(\lambda_{\nu}
G^{\nu})G^{\mu} \eeq

When the Poisson parentheses of these secondary constraints are
taken with the Hamiltonian, we get new relations which evaluate
$c_{\mu}$ and $d_{\mu}$.  Thus the system is closed and we do not
get new constraints.

Now we study the different classes our constraints may belong. To
be a first class constraint  that constraint should have vanishing
Poisson parentheses  with every other constraint, i.e. for every
$i$ and  $j$,

 \be \{\theta_{i},\theta_{j}\}\equiv 0 .\label{popow} \ee

If a single parenthesis is different from zero, we get second
class constraints. Our constraints turn out to second class.

We form the Faddeev Popov determinant by taking the determinant of
the matrix formed by second class constraints.

\be \Delta_{F}=|det\{\theta_{i},\theta_{j}\}|^{1/2} = g_{\mu
\nu}G^2 + 2G_{\mu}G_{\nu}\ee

We can write this result in the partition function formalism.

 \be Z =\int D\pi D\chi \delta(\theta_{i})\Delta_F  exp \left(-i\int(\dot {\chi } \pi-H_c)\right) \ee

Here $\chi$ is the generic notation for  all the fields and $\pi$
represents all the momenta.  If we integrate over all the momenta,
we get

 \be Z = \int D\overline{\psi} D\psi DG_{\mu} D \lambda_{\mu}D\overline{w}^{\mu}
 D w_{\mu} exp \left(i\int L_{eff}d^4 x \right). \ee

By using ghost fields, the effective lagrangian can be written as

 \beq L_{eff}=\overline{\psi}\left(i\partial \!\!\!/ -ieg\partial
  \!\!\!/ g^{-1}+e\left[ G\!\!\!/+\lambda \!\!\!/
  \right]-m\right)\psi -e^4\lambda_{\mu}G^{\mu}G^{2}-L_{ghost}
  \label{etkinlagran}   \eeq

where $L_{ghost}=\overline{w}^{\mu}(G^2 +
2w_{\lambda}G^{\lambda})w_{\mu}$.  If we take an integral over the
spinor fields the effective action is written as

 \beq S_{eff}=Tr\ln\left[i\partial \!\!\!/-e(ig\partial \!\!\!/
  g^{-1}-G_{\mu}\gamma^{\mu}-\lambda_{\mu}\gamma^{\mu})+m\right]
  \nonumber \\
  -\int dx^4 \left(e^{4}\lambda_{\mu}G^{\mu}\left(G_{\lambda}G^{\lambda}\right)+L_{ghost}\right).\label{intlagran} \eeq

At this point we redefine our fields as

 \beq A_{\mu}&=&-ig\partial_{\mu} g^{-1}+ G_{\mu}+ \lambda_{\mu} \nonumber \\
  F_{\mu}&=&\lambda_{\mu}-G_{\mu} \hspace{2.2cm} \nonumber \\
  J_{\mu}&=&G_{\mu}+\lambda_{\mu}+2g\partial_{\mu} g^{-1}. \label{donusum} \eeq

Using the inverse transformations

 \beq
 ig\partial_{\mu}g^{-1}&=& \frac{2J_{\mu}-2A_{\mu}}{6} \nonumber \\
 G_{\mu}&=& \frac{2A_{\mu}-3F_{\mu}+J_{\mu}}{6} \nonumber \\
 \lambda_{\mu}&=&\frac{2A_{\mu}+3F_{\mu}+J_{\mu}}{6}  \eeq

we write the effective action as

 \be S_{eff}=Tr\ln\left(i\partial \!\!\!/+eA\!\!\!/+m\right)
 + \int dx^4
 \left[e^4\left(A_{\mu}A^{\mu}A_{\lambda}A^{\lambda}\right)+\mbox{other
 terms} \right]
  \label{etkin}.\ee

Here we see that only the $ A_{\mu} $ field would appear
quadratically upon expanding this action.  All the other fields
are raised to third or higher powers. We first take the first
derivatives of the effective action with respect to the fields and
set these expressions equal to zero to kill the tadpoles. The
vacuum expectation values of all the vector fields are zero, also
as dictated by Lorentz invariance. We then take second derivatives
to calculate the propagators, using the zero values wherever these
fields appear. For the $A_{\mu}$ field we get

\beq
\frac{\partial^2 S_{eff}}{\partial A_{\mu}\partial
A_{\nu}}=-\left[\frac{g^2}{(2\pi)^4}\right]tr\int
\frac{\gamma_{\mu}(p\!\!\!/+m)\gamma_{\nu}(p\!\!\!/+q\!\!\!/+m)}{(p^2-m^2)\left[(p+q)^2-m^2\right]}d^4p
\nonumber \\
\nonumber \\
=-(g^2/3\pi^2)(q_{\mu}q_{\nu}-g_{\mu\nu}q^2)\left[\frac{1}{\epsilon}+
\mbox{finite part})\right] \eeq

We do not get finite expressions for the propagators of the other
fields.

 \beq
 \frac{\partial^2 S_{eff}}{\partial F_{\mu}\partial F_{\nu}}\mid_{F_{\mu}=0}&=&0, \hspace{15mm}
 \frac{\partial^2 S_{eff}}{\partial J_{\mu}\partial A_{\nu}}=0 \nonumber\\
 \frac{\partial^2 S_{eff}}{\partial J_{\mu}\partial J_{\nu}}\mid_{J_{\mu}=0}&=&0, \hspace{15mm}
 \frac{\partial^2 S_{eff}}{\partial F_{\mu}\partial A_{\nu}}=0 \eeq

Thus all the fields, except $A^{\mu}$ decouple from the model.
This is also true for the ghost fields. We can write the last
expression as

 \be S_{eff}'=Tr\ln\left(i\partial \!\!\! /+ eA\!\!\! /+m
  \right)+\int dx^4 e^4\left( A_{\mu}A^{\mu}A_{\lambda}A^{\lambda}\right)
  \label{loop}\ee

At this point we expand the logarithm term to write the vector
field kinetic term, being the usual vector field expression.  We
will use the usual massless vector field propagator for the
$A_{\mu}$ field assuming we can invert this expression, if
necessary by introducing a gauge fixing term into the lagrangian.

At this point we want to bring a new interpretation to the old
work.  We will interpret the infinities coming from the $A^{\mu}$
propagator as wave function renormalization. Assuming this
expression can be inverted,  the propagator for this field may be
written as $\frac{\epsilon g^{\mu\nu}}{p^2}$ in the Feynman gauge.
We will use this expression to perform calculations in higher
orders.

\section{Spinor Propagator}

In this section we calculate the above results in higher orders.
To justify our result that no mass is generated for the fermion we
may study the Bethe-Salpeter equation obeyed for this propagator.
The Dyson-Schwinger equation for the spinor propagator is written
as

\begin{equation}
 iA p\!\!\!/ +B = i p\!\!\!/ + 4\pi \epsilon \gamma^{\mu}
 \int {{d^4 q }\over {( i A q\!\!\!/ +B)(p-q)^2}} \gamma_{\mu} .
 \label{DSpropagator}\end{equation}

Here $iA p\!\!\!/ +B$ is the dressed fermion propagator. We use
the one loop result for the scalar propagator. After rationalizing
the denominator, we can take the trace of this expression over the
$\gamma$ matrices to give us

\begin{equation}
 B= 16\pi \epsilon \int  d^4 q {{B}\over {(A^2 q^2 +B^2)(p-q)^2}}.
\end{equation}

The angular integral on the right hand side can be performed to
give

\begin{equation}
B= 16\pi \epsilon \left[ \int_{0}^{p^2} dq^2 {{q^2 B}\over{
p^2(A^2 q^2+ B^2)}}+ \int_{p^2}^{\infty} dq^2  {{ B}\over{ (A^2
q^2+ B^2)}}\right].
\end{equation}

If we differentiate this expression with respect to $p^2$ on both
sides, we get

\begin{equation}
{{dB}\over{dp^2}}= -16\pi \epsilon \left[ \int_{0}^{p^2} dq^2
{{q^2 B}\over{ (p^2)^2(A^2 q^2+ B^2)}}\right].
\end{equation}

This integral is clearly finite.  We get zero for the right hand
side as $\epsilon$ goes to zero. Since mass is equal to $m$ in the
free case we get this constant equal to $m$. This choice satisfies
eq.(\ref{DSpropagator}).

The similar argument can be used to show that $A$ is the dressed
spinor propagator is a constant. We multiply
eq.(\ref{DSpropagator}) by $p\!\!\!/$ and then take the trace over
the spinor indices. We end up with

\begin{equation}
p^2 A = p^2 - 8\pi \epsilon \left[ \int_{0}^{p^2} dq^2 \left(
{{(q^4) A}\over{p^2 (A^2 q^2+ B^2)}}+ \int_{p^2}^{\infty} dq^2
{{p^2 A}\over{ (A^2 q^2+ B^2)}}\right)\right].
\end{equation}

We divide both sides by $p^2$ and differentiate with respect to
$p^2$.  The end result

\begin{equation}
{{dA}\over{dp^2}} = 16\pi \epsilon  \int_{0}^{p^2} dq^2 \left(
{{(q^4) A}\over{ (p^2)^3(A^2 q^2+ B^2)}}\right).
\end{equation}

shows that $A$ is a constant as $\epsilon$ goes to zero. Since the
integral is finite, it equals unity for the free case, we take
$A=1$.

\section{Other Processes}

In this section we will try to analyze the contribution of the
higher order diagrams to our basic terms, i.e. to the  three and
four point functions.  We will use only the contributions of the
ladder diagrams, anticipating the result we would have if we had
an inner symmetry index N to justify an ${{1}\over{N}}$ expansion.

Our model is not a gauge invariant model. The gauge invariance,
which may be present in the original model via the $g$ term, is
broken by the new quartic term for the $A_{\mu}$ fields. If this
term were absent, we could fix the gauge, say by imposing the
Lorentz gauge, we would bring new, this time propagating ghost
fields. As it can be shown easily, the additional contribution of
these to the vector field propagator would again be of the
transverse type. For the vector field propagator, we assume that
we can invert  the inverse propagator by, if necessary adding the
necessary term by hand, as an additional constraint. We note that
the $\epsilon$, which is in the denominator in the inverse
propagator, is brought to the denominator in the propagator.

We will first study the four spinor diagrams.  We do not have four
spinor coupling in our Lagrangian.  We need vector particles
coupling to spinors to obtain this expression, which necessitates
the use of vector propagators as internal lines.  Since each
vector propagator contains an $\epsilon$ contribution, the four
spinor process goes to zero as $\epsilon$.  When we go beyond tree
diagrams, we need at least two vector propagators to end up with
two spinors, which means extra powers of  $\epsilon$.  This is
true for all the higher order processes.

We can justify our claims also by writing the Bethe-Salpeter
equations for this process. The Bethe-Salpeter equation for the
four spinor interaction reads as

\begin{equation}
G^{(2)}(p,q;P) = G^{(2)}_{0} (p,q;P) + {{1}\over{(2\pi)^8 }} \int
d^4 p' d^4 q' G^{(2)}_{0} (p,p';P) K(p',q';P) G^{(2)} (q',q;P).
\end{equation}

Here $G^{(2)}_{0} (p,q;P)$ is two non-crossing spinor lines, $
G^{(2)} (p,q;P) $ is the proper four point function. $K$ is the
Bethe-Salpeter kernel.

 The four spinor kernel in this expression is at least to the
first power in $\epsilon$.  We can use the quenched ladder
approximation \cite{Miransky}, where the kernel is seperated to a
vector contribution which is limited to bare propagator only,
connected to the proper kernel with two spinor legs. The proper
kernel is of order $\epsilon$.  The contribution of a vector and
two spinor propagators is finite, recalling that the vector
propagator has an extra $\epsilon$.  The total  result is of order
$\epsilon$ for the ${\overline{\psi}}\psi { \overline{\psi}} \psi$
vertex.  This result shows that that the spinor-spinor scattering
process is vanishes as $\epsilon $ goes to zero.   This result
will be important when we study the $ < {\overline{\psi}} \psi
A_{\mu} > $ vertex.

We note that  the $ < {\overline{\psi}} \psi A_{\mu}> $ vertex is
finite, hence does not need any infinite regularization.  The
first correction to the vertex is the one loop diagram.  Here we
have two spinor and one vector propagators.  The divergence coming
from the loop momentum integration is cancelled by the $\epsilon$
factor coming from the vector propagator.  Naively the higher
diagrams do not change this result, since each momentum
integration is accompanied by an $\epsilon$ term.

To establish this result more firmly we write the Dyson-Schwinger
equation for this vertex. Here we use the result of the four
spinor diagram.  Three point function has the vector particle
going to two spinors which go into the four spinor kernel.  The
loop integration brings a divergence which is cancelled by the
$\epsilon$ coming from the kernel.  The end result is the finite
renormalization of the three point vertex.

We see that we do not need an infinite renormalization for the
four $A_{\mu}$ vertex. The first correction to the tree digram is
the box diagram with four spinor propagators.  This diagram in
spinor electrodynamics is known to be finite
\cite{feynman,karplus,ward}; hence the coupling constant for this
process does not {\it{run}}. The finite contribution of this
diagram just renormalizes the coupling constant by a finite
amount. There are no  infinities for this function coming from
higher orders.   The two loop diagrams contains an $A_{\mu}$
propagator, making this diagram finite.  The three-loop diagram
contains eight spinor and two vector propagators, which are linear
in $\epsilon$.  Higher orders also do not give infinite
contributions; so, the sole coupling constant of the model does
not need infinite renormalization.

\section{Conclusion}

As a result of the arguments in the earlier sections, we can
construct a model where the composites can scatter from each
other, whereas the process whose sole result is the scattering of
the constituent spinor fields from each other vanish as the
cut-off is removed. The scattering of the composite fields from
each other will be a finite expression. There is also a
tree-diagram process where the spinor scatters from a composite
particle, a Compton-like scattering, with a finite cross-section.
This diagram can be written in the other channel, which can be
interpreted as spinor production out of vector particles.  There
are also processes where a single spinor scatters with a vector
composite and creating additional vector particles in the tree
approximation. Here we have to exclude any internal vector lines
as a rule, which makes us use tree diagrams only. There may be
insertions to vacuum fluctuations of a spinor field, but these
just add finite contributions to the propagator, since the
presence of vector propagators make these loop diagrams at most
finite. Whenever the composites do not take part as incoming or
outgoing particles, the cross section goes to zero.

The creation of composite particles, out of two incoming
composites is finite, if the outcome is an even number of
composites.  The creation of an odd number of composites is
forbidden by the Furry theorem in the one loop calculation. Any
loop diagram which creates spinor particles as a result of the
interaction of two vector composites vanishes as the cut-off is
removed.  The lowest of these, which creates two spinors,  are the
triangle diagram made out of spinors, or a box diagram, made out
of three spinors and one vector particle. The first expression is
zero and the amplitude vanishes due to the presence of the vector
propagator in the latter case. The diagram which involves the
production of four spinors involves two vector propagators, giving
also zero cross-section, since it is proportional to a power of
$\epsilon$.

We can also have scattering processes where two vector particles
go to an even number of vector particles.  In the one loop
approximation all these diagrams give finite results, like the
case in the standard electrodynamics. Since going to an odd number
of vector particles is forbidden by the Furry theorem, we can also
argue that vector $\phi$ particles can go to an even number of
vector particles only. This assertion is easily checked by
diagrammatic analysis.

As a result of our calculations we find a model which is trivial
for the constituent spinor fields, aside from Compton scattering
in first order only, whereas finite results are obtained for the
scattering of the composite vector particles.

The processes where a single spinor particle giving scattering
with a composite particle and giving rise to additional vector
particles are allowed. Here we have to  exclude any internal
vector lines as a rule, which makes us stick to tree diagrams
only. There may be insertions to a vacuum fluctuations of a spinor
which are not zero, but these contributions are finite.

An addition to the model is to add a constituent complex scalar
field to the model, doing just the complementary thing to the work
of Bardeen et al. \cite {bardeen1, bardeen2}.  Since we already
have a composite vector field, all we can add is a scalar field
which has its kinetic term in the lagrangian. Work on this model
is going on.  We just want to make the remark that it does not
seem to be feasible to generate the gauge interaction of the
scalar field, the $<\phi \partial_{\mu} \phi A^{\mu}> $ term,
using the triangular spinor diagram, due to the Furry theorem.  A
triangular loop wih one $\gamma^{\mu}$ insertion will be zero. A
way out may be introducing a pseudovector composite particle
instead of a vector one in the first place. Another way out would
be to give internal structure to fields, say $U(n)$ symmetry.

 The
$<A_{\mu} A^{\mu} \phi^2>$ can be generated by the spinor box
diagram, thus making contact between the scalar and the vector
particles in the theory, although such an interaction does not
exist in the original lagrangian. Such a term may break the gauge
invariance, though. We still hope a sense can be made out of this
model which may be complementary to the gauged version of the
scalar case. Our model is a toy model. We could not find a
physical system that is effectively described by it.

\vspace{10mm} {\bf{Acknowledgement}}: We thank Bekir Can
L\"{u}tf\"{u}oglu , Kayhan \"{U}lker, and an unknown
correspondence for discussions and both scientific and technical
assistance throughout this work. The work of M.H. is also
supported by TUBA, the Academy of Sciences of Turkey. This work is
also supported by TUBITAK, the Scientific and Technological
Council of Turkey.

\end{document}